\documentclass[11pt]{article}
\usepackage{geometry}                
\usepackage{graphicx}
\usepackage{amssymb}
\usepackage{epstopdf}
\usepackage{amsmath}
\usepackage{natbib}
\usepackage{hyperref}

\providecommand{\keywords}[1]{\textbf{\textit{Keywords:}} #1}

\title{A dynamic, signals-based reinterpretation of microeconomic theory}
\author{Sarang Shah\thanks{PhD Candidate, University of California, Berkeley. sarang.shah@berkeley.edu}}

\date{\today}                                           

\begin{document}
\maketitle

\section*{Abstract}

Economics has long been a science of static equilibria, in which time is a second-order rather than first-order concern. Without time, economic modelers may neglect or obscure the role of time-dependent phenomena, e.g. path-dependency, and limit their ability to compare agnostically the model results with empirical observations. In this article, I outline a dynamic, signals-based recipe for building microeconomic models from traditional static models. I demonstrate this recipe using a classic ``desert island'' Robinson Crusoe (RC) model of consumption. Starting from a classic static derivation, I then move to a dynamic view, using the utility function as a generator of force on consumption. Finally, I show that the resulting dynamic model may be expressed in Lagrangian and Hamiltonian terms. I conclude by suggesting a recipe for scientific iteration using these alternate mechanical formulations, and the alternative explanations these dynamic models may suggest compared to employing a static approach to modeling.

Word count: 7,938 words

\keywords{theory, models, microeconomics, dynamics, mechanics}

\newpage
\tableofcontents

\section{Introduction}



Economics has long been a science of static equilibria, and has generally not been situated in time. In the simplest models of exchange, economists construct models to ascertain an ultimate distribution of endowments through rational exchange, such that it is neither important how quickly nor along which exchange paths one arrives at this static equilibrium. Time, in this view, often enters in the form of comparative statics, examining the passage from one static equilibrium to another over time, after the climactic action of forming a static equilibrium has already taken place. Or time may enter into the determination of the equilibrium itself, through some version of Walrasian \emph{tâtonnement}. Yet, this groping toward equilibrium is not meant to be taken at face value as a description of an observed dynamic process but as a hypothetical, an imagined as-if clearing mechanism to arrive at the final exchange distribution. In either integration of time into static models, we do not directly observe these dynamic changes over time, nor do we bring our models into contact with these observations of dynamics apart from an observation of the initial and the final states.

What if, however, we took economic measurements and observations over time, i.e. signals, more seriously as the primitive unit of our models? Rather than treat notional preferences, revealed through actions, as the primitive unit of economic model-making, we could instead use observations of real life economic signals, of quantities purchased and consumed, prices, taxes and remittances, etc., over time.

Indeed, much of our current empirical understandings of preferences, and therefore indifference curves, utility functions, and supply and demand curves, originate from some set of observations over time. Sometimes these observations are collected in the context of a laboratory experiment, such as through subjecting behavioral subjects to binary questionnaires of the kind envisioned by \cite{Rubinstein_2012}. Other times, our observations come from some rich contextual market environment, such as the record of sales made at a grocery store\footnote{See \cite{Klenow_Malin_2010} for a review of recent work on these data sets.}. Either way, experimental or observed, a context has been created through laws, rules, and other institutions\footnote{See \cite{Veblen_1969} on the role of institutional contexts in shaping preferences.}, agents are selected, and some experimental construct, either created by us or ``natural,'' generates a set of signals we can then unpack. Many markets worthy of our study and understanding are signal-producing social machines.


I am hardly the first to prescribe a dynamic formulation of economics \citep{Estola_2017, Glotzl_Glotzl_Richters_2019, McLaren_2016, Chen_2010, Aubin_1997, Cass_Shell_1976, Sato_1981}. Many have developed a version of microeconomic dynamics in the context of statistical finance or econophysics \citep{Haven_Khrennikov_2017, Frieden_Hawkins_DAnna_2007, Sato_Ramachandran_2014}.

There are several strengths to a dynamic, signals-first approach to economic modeling. The introduction of time by focusing on economic signals allows us to model specific time-based features through the construction of dynamics. Phenomena over time may experience path-dependency and hysteresis for instance, which themselves may be aspects of irreversibility. Our constructed dynamics may be decomposed into those parts that are dynamically conservative and reversible, and those parts which are dissipative and irreversible, with each part bearing interpretive significance. The importance of sequencing may be revealed through these dynamics, where interventions or actions either on the system or of the system on itself perturb or change the system’s informational environment. Dynamics, unlike statics, may formally model probabilistic or stochastic observables through ``canonical quantization,'' which would allow us to formally invite uncertainty back into our economic models.\footnote{Though beyond the scope of this essay, it is important to note here that while all of the above, but especially canonical quantization, may bear some superficial resemblance to the mathematics of quantum theory, this is largely due to the integration of both economic dynamics and quantum theory into an emerging information-theoretic paradigm, an integration available to all scientific studies of measurable observables over time. For a brief introduction to information theory and its applications, see \cite{pierce1980introduction}}

A dynamic approach may also follow the ``agnostic'' approach to empirical causal inference that has come to pervade contemporary empirical economics.\footnote{See \citep{Aronow_Miller_2019} for an in-depth exploration of this agnostic view of empirical economics} Rather than model systems \emph{ex ante} with \emph{a priori} assumptions, much post-credibility revolution empirical work focuses on tactically using statistical models to draw an inference of causation embedded in some rich qualitative context. Likewise, by treating economic signals as the primitive epistemic unit, we can reduce our reliance on a priori assumptions, such as transitivity, completeness, etc., to derive an initial explanatory model of our system from dynamic signals alone, and allow ourselves room to ground whatever model we construct in these observations rather than try to fit them into an existing rational-agent, utility-maximizing model.\footnote{Empirical and theoretical microeconomics, contrary to popular conception outside of economics, remains a thriving, even burgeoning field of study. A rise in computational power and innovations in experimental designs have given new life to the field once thought dying post-Great Financial Crisis (for a brief review, see \cite{Einav_Levin_2015}). These studies take preferences seriously, while also integrating findings from behavioral economics by hammering kinky curves into well-behaved ones. By defining an efficiency index, classical microeconomists may, under certain constraints, rely on revealed preferences to define utility curves while also making comparison with other empirical observations (see e.g. \cite{Polisson_2020}. Nevertheless, this research program still relies on finding some \emph{mens rea} of individual preferences and adopting strong assumptions regarding how preferences translate into actions.} An agnostic approach to economic signals may indeed lead us to deriving a rational-agent, utility-maximizing view of the observed system in question, but it may also suggest other modes of explanation that fit the observations just as well if not better. Such versatility is crucial when traditional assumptions about economic behavior just cannot be sustained.

There are however some weaknesses in these prior dynamic approaches, despite their impressive formalizations and applications. Prior dynamic approaches to modeling start with assumptions about behavior and agents, eschewing a pragmatic, agnostic view of what produces economic signals. Forces on consumption, for example, are taken to be indicative of some behavioral or cognitive process that in a reductionist way wait for a clever empiricist to seek them out. Prices are seen as negotiated through a \emph{tâtonnement} process, one that could be found through observation. To believe that there are mechanisms that may be found and observed behind dynamics is not necessarily wrong; however, I argue that to believe such an observable driver of dynamics is often unnecessary for two reasons. First, a model need not be reducible to behavior, cognitive processes, or the push and pull of physical or biological forces, to be useful. We may pragmatically model many things, such as the impact of policy interventions, without having to tease apart the forces we observe. Indeed, many of these forces may be irreducible, an emergent result of some activity beyond our ability to discern empirically. Second, we need not believe a force on a signal exists in order to posit one for the sake of model-building. When we wish to use a model to calculate or probe the movement of signals over time, we can hold out concepts of inertia, force, etc. not as real (and therefore open to empirical observation), but as convenient and pragmatic accounting tools for our purposes. A pragmatic view of signal dynamics over time opens our toolkit to that which works or that which we creatively intuit, and from which we may, if we so choose, imagine a wider range of putative mechanisms to be empirically observed.

Economic models have frequently been accused of being out-of-touch with empirical reality\footnote{See \cite{Spiegler_2015} for an excellent exploration of the perils of abstraction in economic models and how models may be rehabilitated through a hermeneutic, interpretive turn.}. I do not argue that we should drift further from empirical observations; rather, I suggest that interpretation and pragmatic model-building are two complementary, yet separate tasks.

These prior approaches to dynamics also still hold the discovery of equilibria as their ultimate objective. It is true that equilibria may emerge from dynamic descriptions, and as described above, a dynamic approach may encompass and extend a classical statics-based approach by describing how an equilibria is arrived at and the interesting dynamic features that may emerge along the way. Yet, one of the advantages of a dynamic approach that a statics approach could never achieve is to describe firmly non-equilibrium systems. In fact, most systems do not achieve equilibrium, as there are very few systems (outside of physics, chemistry, and certain engineering applications) for which the system cut between experimental apparatus and experiment is sufficiently impermeable to perturbations and exchange from outside the system. At best, a dynamic model of an economic system may not show a true equilibrium so much as a stationary state for some period of time, and may even indicate to us some half-life of decay. More generally, a dynamic model may simply show dissipation, emergent structures, and interaction for some limited period of time. 


Finally, few proposed dynamic approaches have offered both a recipe for cumulative scientific inquiry and a pathway for transforming current static microeconomic models into dynamic ones. As mentioned above, this approach can accommodate and integrate models already built to describe a wide variety of markets. As such, from a sociology of knowledge perspective, finding some pathway from static to dynamic models offers some benefits. First, utility functions and models created using the classical approach can be “plugged into” this framework, yielding a useful starting point for further model-building and inquiry, even if it is ultimately discarded. Second, those who have spent their careers developing these models may find some simple formal adaptations to their existing approaches to bring them into a dynamic framework. While the oft-repeated quip that scientific revolutions proceeds at the pace of deaths and retirements, it is also the case that scientific revolutions are made much more palatable by integrating and repurposing the knowledge and human materiel that has come before. This move toward signals takes the empirical and experimental contexts for eliciting preference information and situates it as one of many contexts through which some economic signal may be generated. In this case, those signals are preferences, which through a series of assumptions and mathematical \emph{ansätze}, such as those of canonical choice and decision theories, yield much of traditional microeconomics. 

In this paper, I offer one such pathway. I use a simple pedagogical example to outline a recipe for converting static microeconomic models into dynamic signals-generating models, and how one can engage iterate scientifically using these models. Shifting from treating ever-elusive preferences as our primitive units to signals composed of economic observations opens ourselves to an empirical and theoretical agenda with many benefits. Rather than outline a competing alternative to economic theory-building, I show that the approach outlined in this paper can be adopted by anyone engaged in theory-building, even those who believe the assumptions of methodological individualism, choice and decision theories, utility and preferences, etc. are indeed suitable for that which they are seeking to explain and model. Moreover, the pathway described below illuminates how a theory of signals lies \emph{sub rosa} in the foundations of static microeconomic theory.\footnote{In this, I follow Fisher and Pareto in holding a theory of signals could serve a research program examining how some notion of utility could drive change in measurable observables over time. See \cite{fisher1892mathematical} and \cite{pareto1896ecrits}.}

I show how this linkage can be made by playing with a classic ``desert island'' Robinson Crusoe (RC) model. These models, well known to beginner students in economics, have long been used as pedagogical tools to demonstrate key features of consumption and exchange. I start with the classic statics-based description of RC consumption. I then move to a dynamic view, using the utility function as a generator of force on consumption. Finally, I show that the resulting dynamic model may be expressed in Lagrangian and Hamiltonian mechanics terms, free of any expression of utility, preferences, or agents. I also repeat these steps for the example of exchange in the Appendix. I finish by suggesting alternate approaches to building these models and alternative explanations these models may suggest had I not started from rational agents maximizing their individual utility.

I pay particular attention to constructing the Lagrangian and Hamiltonian formulations of dynamics because I believe both open our model-building approach into a vast, versatile, richly-textured world of applied mechanics. By connecting our island of microeconomics to this mainland, we may access a much broader armory of formal mathematics and theory-building to advance our research agenda. We also bring microeconomics full-circle to its origin point, as a useful, constructive dialogue between social observations and physical modeling methods. Our hope is that the demonstration that follows offers a glimmer of hope of how we may find our way to the farthest shore of dynamics without losing the accumulated wisdom of static microeconomic models.

\section{Robinson Crusoe Model of Consumption}

The Robinson Crusoe models are highly stylized stories used in economics pedagogy to illustrate key features of consumption and exchange while also demonstrating the practical application of the classical microeconomic modeling approach. In the following section, I explore consumption, by first elaborating the static equilibria solution using the traditional approach. Next, I derive a dynamic model starting from the static model, and from those models determine the static ``equilibrium'' solution. Finally, I explore the Lagrangian and Hamiltonian formulation of these dynamics. I repeat the above steps for exchange in the Appendix.

My aim is to fill in the gap between microeconomic theory and the study of economic signals, by starting with the simplest example of how microeconomic model-building is taught to economics, policy, and law students, and how this example could be transformed into a dynamic one, with all the advantages such a transformation entails. I conclude with reflections on how this ``recipe'' for model-building opens up to a general \emph{agnostic} approach to normal science.

\subsection{Consumption}

We start with Robinson Crusoe stranded on a desert island alone, with none of his belongings. He is hungry, and notices the island is abundant with only two foods, bananas ($B$) and coconuts ($C$). The island is so abundant with these fruits that he need not worry about overconsumption. His utility may be described using an utility function in Cobb-Douglas functional form\footnote{As noted in the previous section, the Cobb-Douglas utility function encapsulates many typical assumptions baked into microeconomic theory, namely diminishing marginal utility, as well as transitivity and completeness of preferences.}, where $Q_B$ is the total quantity of bananas consumed, $Q_C$ is the total quantity of coconuts consumed, utility $U$ is his total utility with units $utils$\footnote{What are $utils$?The concept of utility is contested and fraught, even within the mainstream economics discipline. Is utility some measure of pleasure? Or does it strictly correspond with use value? For the purposes of this essay, $utils$ are a unit quantity for utility without a specific interpretation or measurement. As we will see later, the resulting dynamics, as well as their corresponding Lagrangian/Hamiltonian are indifferent to the choice of unit for utility at this stage, and given that this unit ultimately drops out of the dynamic formulation. The dynamic approach may thus possess the advantage of using utility as a placeholder concept without necessarily stipulating a specific interpretation for it.}, $k$ is a proportionality constant with units $utils/(bananas \times coconuts)$.

\begin{equation} \label{eq:utility}
U=kQ_B\cdot Q_C
\end{equation}

Robinson Crusoe possesses a fixed amount of energy $E$ a day; he can only pick so many bananas and coconuts given his energy budget. The energy cost of picking bananas is fixed at $p_B$ and coconuts at $p_C$, each with units $energy/banana$ and $energy/coconut$ respectively. 

Robinson Crusoe is assumed to spend his entire energy budget to maximize his utility. In other words, Robinson Crusoe spends his entire budget $E$ on picking bananas and coconuts: 
\begin{equation} \label{eq:budget}
E=p_BQ_B+p_CQ_C
\end{equation}

We now have the classic budget constraint \eqref{eq:budget} and utility function \eqref{eq:utility} of traditional microeconomics. Let us continue to the fairly straightforward static determination of the bundle of bananas and coconuts consumed by Robinson Crusoe.

\subsubsection{Static Consumption}

If we were working in a purely static context in which the consumer, with knowledge of their own utility function, immediately chooses the bundle of goods $B$ and $C$ that maximizes their own utility, we would get the following consumption result.\footnote{Typically, a microeconomic modeler would employ a Lagrangian multiplier approach to maximize utility subject to the budget constraint above, but since our budget and utility functions are so simple, we will use a straightforward identification of a local maximum of the utility function as follows.}

\begin{enumerate}
\item First, let us rewrite the budget constraint to solve for $Q_C$:
\begin{equation}
Q_C = \left(1/p_C\right)\left(E-p_BQ_B\right)
\end{equation}
\item Next, let us express total utility $U$ as solely a function of $Q_B$ given the budget constraint above:
\begin{equation}
\begin{split}
U &= k Q_B Q_C \\
&= k \frac{Q_B}{p_C}\left(E-p_B Q_B\right) \\
&= Q_B \left(\frac{k}{p_C}E\right) - Q_B^2\left(k \frac{p_B}{p_C}\right)
\end{split}
\end{equation}
\item To identify the local maximum of $U$, we find the point $Q_B^*$ where $\partial U/\partial Q_B=0$\footnote{This reduced Cobb-Douglas utility function satisfies the second-order sufficiency condition for the existence of a local maximum: $d^2 u/d\dot{Q_B}^2 = -2k\frac{p_B}{p_C}<0$}:
\begin{equation}
\begin{split}
\partial/\partial Q_B \left[Q_B \left(\frac{k}{p_C}E\right) - Q_B^2\left(k \frac{p_B}{p_C}\right)\right] &= 0 \\
\frac{k}{p_C}E-2p_B\frac{k}{p_C}Q_B &= 0
\end{split}
\end{equation}
\item Solving for $Q_B$ we find:
\begin{equation}\label{eq:totalbananas}
Q_B^* = \frac{E}{2p_B}
\end{equation}
\item Finally, we plug $Q_B^*$ into our budget equation to find $Q_C^*=\frac{E}{2p_C}$.
\end{enumerate}

Robinson Crusoe would therefore consume, in total, $Q_B^* = \frac{E}{2p_B}$ bananas and $Q_C^*=\frac{E}{2p_C}$, which maximizes his utility from bananas and coconuts while exhausting his energy budget.

From the assumption that a rational consumer would seek to maximize their utility through their selection of consumption bundle, we would usually stop here. If there is any change over time, we would model it through comparative statics, analyzing which parametric changes would lead to a new static equilibrium, and then positing ways that a consumer may move from one equilibrium to another over time. In the next section, we will consider not what Robinson Crusoe's total consumption would be, but how a drive to optimize his utility leads him to this total consumption amount over time.

\subsubsection{Dynamic Consumption}

Let us consider the following. Robinson Crusoe does not pick and consume bananas and coconuts all at once. Rather, he picks some bananas and some coconuts at any given moment in time. He still seeks to maximize his utility, and to use up his energy budget. But his fruit-picking and consumption are strategic over time. As in the static case, he is driven to pick and consume according to some internal utility function. This utility function exerts, in a sense, a \emph{force}\footnote{To aid in developing these dynamic models, I follow \cite{Estola_2017} and \cite{Glotzl_Glotzl_Richters_2019} in introducing the concept of a generalized force as a constructed quantity corresponding to some rate of change.\footnote{\cite{fisher1892mathematical} and \cite{pareto1896ecrits} also introduced the notion of force as corresponding to marginal utility.} Unlike Newtonian force, i.e. force that is the product of mass and acceleration per Newton's second law $F=m\cdot a$, our force is simply the product of some inertial quantity and a rate of change in some base quantity (either a quantity consumed or a rate of quantity consumed) $f=m\cdot \dot{Q}$.} on Robinson Crusoe's consumption that drives him toward maximizing it. Robinson Crusoe does not know his own utility function, so he cannot simply plan to pick and consume the optimal amount of bananas and coconuts all at once, or he may be limited in how much he can consume at a given moment, but he can follow this internal compass to pick some amount of bananas and coconuts at any given moment that increases his utility toward a local maximum.

Given the budget constraint, that Robinson Crusoe chooses to pick and consume bananas in proportion to the degree to which a marginal change in doing so changes his utility $\partial U/\partial Q_B$, with units $utils/bananas$. His instantaneous rate of consumption of bananas $\dot{Q_B}$ is proportional to $\partial U/\partial Q_B$ by some inertial constant $m_B$ with units $(utils \cdot time)/(bananas^2)$\footnote{$time$ is a placeholder for an appropriate measure of time: $s$, $min$, $hr$, etc.}
\begin{equation}\label{eq:2ndlawconsumption}
m_B\dot{Q_B} =\partial U/\partial Q_B \\
\end{equation}

Plugging in our Cobb-Douglas utility function from before, we first compute $\partial U/\partial Q_B$:
\begin{equation}
\partial U/\partial Q_B = \partial/\partial Q_B\left[k/p_C Q_B\left(E-p_BQ_B\right)\right] = k/p_C\left(E-2p_BQ_B\right)
\end{equation}
Instead of setting this partial derivative to $0$, as in the static case, we instead solve the differential equation \eqref{eq:2ndlawconsumption}:
\begin{equation}
\begin{split}
m_B\dot{Q_B} & =\partial U/\partial Q_B \\
&= k/p_C\left(E-2p_BQ_B\right)
\end{split}
\end{equation}

A simple computation yields the following result for $Q_B(t)$:
\begin{equation}
Q_B(t) = b_0e^{-2(k/m_B)(p_B/p_C)t}+\frac{E}{2p_B}
\end{equation}

$b_0$ in this formulation corresponds to the solution to the equation given initial endowment at time $t=0$:
\begin{equation}
b_0 = Q_B(0)-\frac{E}{2p_B}
\end{equation}

Assuming that $Q_B(0) = 0$, which also ensures that Robinson Crusoe does not pick and consume negative bananas, we find the following equation for $Q_B(t)$:
\begin{equation}
Q_B(t)=\frac{E}{2p_B}\left(1-e^{-2(k/m_B)(p_B/p_C)t}\right)
\end{equation}

As $t\to 0$, $e^{-2(k/m_B)(p_B/p_C)t}\to 1$, and as $t\to \infty$, $e^{-2(k/m_B)(p_B/p_C)t}\to 0$, which means that at time $t=0$, $Q(0)=0$ and at time $t\to \infty$, $Q(t) = \frac{E}{2p_B}$. Starting from zero bananas, Robinson Crusoe picks and consumes bananas in such quantities over time until he has consumed $\frac{E}{2p_B}$ bananas as $t\to \infty$.

We now have a dynamic equation for bananas consumed over time, which also yields dynamic equation for coconuts consumed over time given the budget constraint \eqref{eq:budget}. We also observe that when we set the constants of our differential equation's solution to the proper initial conditions, we have a sensible trajectory of picking and consumption that leads to our previously computed static equilibrium solution over time \eqref{eq:totalbananas}. The last thing to note are the units of $k/m_B$. Recall that the units of $k$ are $utils/(bananas\cdot coconuts)$ while the units of $m_B$ are $utils/bananas^2$. The units of $k/m_B$ are therefore $bananas/coconuts$ while the units of $p_B/p_C$ are $coconuts/bananas$, thus leaving the coefficient of exponential increase unitless, thus independent of the choice of unit $utils$.

What do we gain from this dynamic formulation? We have a dynamic function $Q(t)$ which plots quantities picked and consumed over time. Such a signal could be observed and plotted against this model signal, and the clear comparison between the model and observed signals yields greater degrees of freedom than a static view in modifying our model. Second, the first-order force $f=m\dot{Q}$ need not necessarily be presumed as the marginal utility for marginal quantity consumed. In this case, we specifically tied the dynamic signal $Q_B$ to an interpretation of utility as a driver of consumption. But an alternate interpretation may suggest some other driver for the forces on signals. Finally, this dynamic formulation permits us to draw further inferences about the system that a static formulation would obscure. We draw some of these inferences in the following section on the Lagrangian and Hamiltonian formulation.

\subsubsection{Lagrangian and Hamiltonian}

The Lagrangian and Hamiltonian formulations are means of expressing the dynamics in such form that the \emph{action} of the dynamics is minimized. The action of a dynamic signal is a quantity accumulated over time. The action is defined then as this quantity that when minimized produces the observed or putative dynamic trajectory. The Lagrangian is a derivative quantity of this action, and the Hamiltonian is an alternate formulation of the Lagrangian in a different set of canonical coordinates. The mathematical foundations of Lagrangian and Hamiltonian mechanics are beyond the scope of this essay. What we will instead use are heuristic recipes for producing the Lagrangian and Hamiltonian that permit us to skip integral calculus in lieu of a more tractable (yet equivalent) differential calculus. As such, consider the celebrated Euler-Lagrange equations:
\begin{equation}\label{eq:eulerlagrange}
\frac{d}{dt}\left(\frac{\partial \mathcal{L}}{\partial \dot{Q}}\right)-\frac{\partial \mathcal{L}}{\partial \dot{Q}} = 0
\end{equation}

Whatever Lagrangian $\mathcal{L}$ we devise must satisfy \eqref{eq:eulerlagrange} for dynamic signal$Q(t)$ when under a \emph{conservative} force. One can intuit the Lagrangian with the following decomposition $\mathcal{L} = T-V$, where $T$ is a first-order kinetic term usually of the form $T=\frac{1}{2}m\dot{Q}^2$, and $V$ is a potential usually of the form $V = -\int F dQ$ when the force $F$ is a conservative force.

Alas, we immediately encounter two serious problems in employing this heuristic. First, our intuition only works for when the posited force acts on second-order time derivatives (e.g. acceleration). In our prior formulation, our force $f=m\dot{Q}=\partial U/\partial Q$ acts at the first-order on time. To input our force into $V$ above, we must take the derivative with respect to time of $f$ to produce $F \equiv \dot{f}$. Taking the derivative with respect to time on both sides of $f=m\dot{Q}$ yields$\dot{f}=m\ddot{Q}$, since $m$ is constant. We can now use $\dot{f}$ as the force on second-order motion, provided we remain attentive to constraining by initial and boundary conditions in the final resulting dynamic equations.\footnote{As a further exercise, one can show that solving the differential equation $\dot{f} = \frac{d}{dt}\left(\frac{k}{p_C}\left(E-2p_BQ_B\right)\right) = -2k\frac{p_B}{p_C}\dot{Q_B} = m_B\ddot{Q_B}$ yields the same dynamic equations as above.}

Once we find that force $F \equiv \dot{f} =-2k\frac{p_B}{p_C}\dot{Q_B}$, we immediately encounter a second problem. We cannot find $V=-\int F dQ_B$ since we must integrate $dQ_B$ over $\dot{Q_B}$:
\begin{equation}
V = -\int \left[-2k\frac{p_B}{p_C}\dot{Q_B}dQ_B\right]
\end{equation}

The above integration is not exactly integrable. Instead, this integration is path-dependent, meaning the particular trajectory one chooses to integrate over the phase space $(Q_B, dQ_B)$ changes the total integrated value $V$. The presence of path-dependency implies that our force is non-conservative, which means we must adopt a different set of heuristics to find the Lagrangian of our dynamics, one that is suited for dynamics under non-conservative (or \emph{dissipative}) forces such as $F$. Thankfully, D'Alembert's Principle allows us to express the Euler-Lagrange equation for a force with a dissipative component as follows:
\begin{equation}\label{eq:dalembert}
\frac{d}{dt}\left(\frac{\partial\mathcal{L}}{\partial\dot{Q_B}}\right) - \frac{\partial\mathcal{L}}{\partial Q_B} = F_d
\end{equation}

In the above equation, our Lagrangian $\mathcal{L}$ only contains that potential which emerges from the component of forces acting on the system that are conservative and exactly integrable. Since there is no such component $V=0$. We do, however, have a dissipative force, which in this case is the entirety of the force acting on the system: $F_d = -2k\frac{p_B}{p_C}\dot{Q_B}$. From this equation, with $\mathcal{L} = T - V = \frac{1}{2}m\dot{Q}^2$ and $F_d = -2k\frac{p_B}{p_C}\dot{Q_B}$, we quickly recover our original dynamic equation\eqref{eq:2ndlawconsumption} from \eqref{eq:dalembert}:

\begin{equation}
\begin{split}
\frac{d}{dt}\left(\frac{\partial\mathcal{L}}{\partial\dot{Q_B}}\right) - \frac{\partial\mathcal{L}}{\partial Q_B} &= F_d \\
\frac{d}{dt}\left(m_B\dot{Q_B}\right) &= -2k\frac{p_B}{p_C}\dot{Q_B} \\
m_A\ddot{Q_B} &= -2k\frac{p_B}{p_C}\dot{Q_B}
\end{split}
\end{equation}

Conservative forces in physics conserve \emph{energy} and are not path-dependent, while non-conservative forces dissipate \emph{energy} and are path-dependent. In this context, what we are meant to interpret as energy is not clear. Yet, as discussed below, we may get a clearer idea of what energy as action represents in the Hamiltonian formulation. \footnote{\cite{Hands_1992}, in his constructive engagement with \cite{Mirowski_1992}, notes that price is often taken implicitly as a conservative vector field over the scalar potential of compensated demand, the expenditure function. What we have found is that the utility/expenditure function are nothing of the sort, yet the hope that utility would, when integrated, serve as some sort of analogue to the action over time of physics has bedeviled economic theorists since Walras. Instead, what we find is that utility gives rise to a non-conservative, frictional force, which in turn dissipates the energy of an integrated Lagrangian action over time. The scalar potential we do find, $V$, the gradient of which was hoped to be a means of expressing price (or quantity consumed) is simply equal to $0$.}

In any case, the breakdown of force into these two components, conservative and non-conservative, is a useful addition to the consideration of microeconomic signal dynamics. In this case, there was no conservative component to the force acting on the system due to utility. Yet, if the market signal of quantity consumed of some good has a force that can be decomposed into conservative and non-conservative components, then the non-conservative force can be interpreted as a \emph{friction} that drags the signal in an opposing direction proportional to the signal's first-order time derivative. In such a case, the path to equilibrium cannot be instantaneous, and that some path-dependent effect must occur.

Whereas in the static equilibrium scenario, the utility function proportionality coefficient $k$ drops out of the final equilibrium solutions $Q_B^* = \frac{E}{2p_B}$ and $Q_C ^* = \frac{E}{2p_C}$, here it appears to persist in the rate coefficient $k/m_B$ of the exponential term! Given prices $p_B$ and $p_C$, we can arrive at an empirically observable dissipative coefficient $k/m_B$ independent of the point of static equilibrium consumption. This coefficient can even be parametrized into its own model of the frictional component of the overall dynamic of quantity consumed over time. 

This force of friction was a necessary consequence of \eqref{eq:2ndlawconsumption}. This means that even our simple Cobb-Douglas utility function implies some frictional, non-conservative force that pushes consumers seeking to maximize their utility at the margins toward a local maximum (with the relevant coefficient of friction determined or modeled exogenously). Friction, drag, dissipation, and damping are metaphors that have typically been situated at the margins of microeconomic and transaction cost theory\footnote{See Grewal (2024)} as ''corrections`` to the proper, efficient "equilibrium" functioning of a market. In this simple example of consumption, however, friction moves into the center of our interpretation and construction of economic metaphors, and equilibrium vanishes entirely. Rather than arrive at equilibrium, a state of a balance of forces, we instead have a complete dissipation of kinetic energy due to a frictional force. While it behooves us not to take the energy and friction metaphors too literally, it does allow us to reconceptualize even the most basic microeconomic setups away from the pursuit of equilibrium solutions.\footnote{Indeed, many equilibria in economics may be described as either frictional dissipation. Where conservative forces exist, such as those constructed in \cite{Estola_2017}, these equilibria may be better described as stationary states for which a constant input of force from outside the system sustains the perceived equilibrium.}


Before we move on, let us briefly derive the alternative Hamiltonian formulation of our consumption dynamics. Again, without dwelling on derivations, the Hamiltonian is the Lagrangian after a Legendre Transformation where the quantity $\pi$ is defined as $\pi\equiv m\dot{Q}$. If $\mathcal{L}=T-V$ then the canonical Hamiltonian is $\mathcal{H}=T+V$. 

More generally, the Hamiltonian is defined as satisfying the Hamilton-Jacobi equations, produced through a Legendre transformation of the Euler-Lagrange equations, for a given dynamic:
\begin{equation}
\begin{split}
\dot{Q_B} &= \frac{\partial \mathcal{H}}{\partial \pi_B} \\
\dot{\pi_B} + \frac{\partial \mathcal{H}}{\partial Q_B} &= F_d
\end{split}
\end{equation}

Note above that we have chosen to use the D'Alembertian form of the Lagrangian for a system with a dissipative force as outlined earlier. If we take the heuristic that $\mathcal{H}=T+V$, giving us $\mathcal{H}=\frac{\pi_B^2}{2m_B}-\frac{k}{p_B}EQ_B$. Plugging $\mathcal{H}$ into the Hamilton-Jacobi equations above, we find:

\begin{equation}
\dot{Q_B} = \frac{\partial \mathcal{H}}{\partial \pi_B}  = \frac{\pi_B}{m_B}
\end{equation}
\begin{equation}
\begin{split}
\dot{\pi_B} + \frac{\partial \mathcal{H}}{\partial Q_B} &= F_d \\
\dot{\pi_B} &= -2k\frac{p_B}{p_C}\dot{Q_B} \\
m_B \ddot{Q_B} &= -2k\frac{p_B}{p_C}\dot{Q_B}
\end{split}
\end{equation}

As a reminder, $V=0$ since there are no conservative forces acting on the system.The first Hamilton-Jacobi equation gives us back our transformation to $\pi_B$, and the second yields the equations of motion once more for the consumption of bananas $B$, $Q_B(t)$.

As with defining our Lagrangian, defining our Hamiltonian is both a creative act (as the Hamiltonian is that which the solution of the Hamiltonian-Jacobi equation gives us our specified dynamic signal) and a very useful tool for probing dynamics. First, the Hamiltonian may represent an \emph{ersatz} total \emph{energy} as it sums a kinetic term $T(\dot{Q})$ and potential term $V(Q)$. $\mathcal{H}$ is time-independent,$d\mathcal{H}/dt=0$, implying that $\mathcal{H}$ is some constant quantity at the systemic level, among the observables composing $\mathcal{H}$, that is conserved over time. \cite{Meyerson_1908} had astutely pointed out that in modeling over time, the identification of regularities always implies some conserved quantity over time. \cite{Veblen_1969} also suggests that whatever appears conserved may itself be a consequence of institutional conditions that have remained constant over time. We may likewise attribute the constancy of the Hamiltonian to fixed institutions.

However elusive this conserved quantity may be, it is always available as a construction of the available observables, whichever ones generate a particular observed set of dynamic signals via the Hamilton-Jacobi equations. In this sense, the Hamiltonian is a powerful tool. In addition to the power of the Lagrangian to separate conservative and non-conservative forces and to permit coordinate and gauge transformations, the Hamiltonian also represents the model as a whole: a scalar function constant over time, yet embedding the couplings between observables necessary to generate their dynamics. 

Since in addition to a Hamiltonian form we have also found a dissipative force $F_d$ is a key divergence from classical views of conservation in static equilibrium theories. Classical theories imply that the sum of utility and expenditure is conserved: the utility of the currency spent equates to the utility gained from that which is bought and consumed. Here, however, some otherwise conserved quantity is lost in the dissipative force $F_d = -2k\frac{p_B}{p_C}\dot{Q_B}$. If $\mathcal{H}=\frac{p_B^2}{2m_B}-\frac{k}{p_B}EQ_B$ represents some \emph{energy} conserved, then there is necessarily less of it due to the drag exerted by $F_d$ on the rate of consumption over time. Perhaps we can interpret this additional quantity lost as a \emph{vigor} of consumption that must necessarily be diminished as the consumer arrives at a level consumption that maximizes their utility. It is not necessary, however, to posit interpretations at this stage. The Hamiltonian formulation offers us a window into an agnostic, non-deterministic approach to modeling, one that takes the contingency of modeling seriously by first allowing us to separate the conserved quantity constructed from our observables from the dissipative force.\footnote{The Hamiltonian approach also us to approach the underlying sources of stochasticity in the economic data we observe. While a full elaboration is beyond the scope of this paper, it is worth noting that the Hamiltonian form is the typical starting point for stochastic and quantized models, usually achieved through ``canonical quantization,'' which treats each observable in the Hamiltonian as an operator from an operator algebra acting on a Hilbert space of "state vectors", with the Hamiltonian itself as the operator evolving state vectors over time. The significance of this approach is two-fold. First, it allows us to work with stochastic dynamic trajectories, while still inferring and intuiting conservation laws, time-dependent effects, and other features that a more deterministic dynamic view would furnish. Second, we would be able to formally account for the epistemological realization that many economic measurable observables are the result of transaction and exchange, or some other interaction, which in principle perturbs the informational environment of the system. For example, price in certain markets is the result of negotiation and transaction, and in principle, the price signal is generated through these actions rather than existing \emph{ex ante} waiting to be discovered. The appearance in observed data of this idea is often a degree of stochasticity and uncertainty. As financial economists have long acknowledged through tools such as the Black-Scholes equation, this stochasticity and uncertainty can indeed be quite significant.} 

For the application of the above steps to the case of exchange, please see \nameref{appendix}.

\subsection{A recipe for normal science?}

Is the Lagrangian/Hamiltonian approach conducive to the practice of science?

Science may be framed as a practice in which a discursive engagement between theory and empirical observation allows us to develop better models, heuristics, and explanations of observed phenomena. We measure the success of a recipe for scientific practice according to whether subscribers of that practice may generate collective, collaborative, intersubjectively-verifiable understanding. A key feature, I believe, in being able to do so is possessing a model of iteration as some scientific ``recipe'' prescribes. I adopt two modes of scientific iteration:
\begin{enumerate}

\item One mode of iteration is to develop a model based on assumptions, intuitions, and prior work, and to then compare that model to a set of observations to determine how that model may be modified or discarded. 
\item The other mode of iteration is to collect a set of observations and to then develop a model to explain it. 

\end{enumerate}
These modes are complementary and alternating. A theory permits us to probe empirical observation further, suggesting other areas in which to direct our scientific gaze. Observations then allow us to correct and expound our theories. Together, theories become more robust, split off into other theories, or are overturned altogether in a Kuhnian paradigm shift. 

I believe that adopting a Lagrangian/Hamiltonian approach to the study of economic signals is an adequate recipe for scientific practice, and is indeed a meta-framework for the practice of choice and decision theorists.\footnote{The Lagrangian/Hamiltonian approach to theorizing dynamics has also had a remarkable degree of success in financial economics, resource economics, the study of population dynamics, control theory, digital signals processing, thermodynamics and statistical mechanics, and quantum theory.} 

To demonstrate this second mode of scientific iteration, let us attempt an interpretation of the dynamics of consumption from the section above as if we were observing it, not constructing it. Let us assume that Robinson Crusoe was indeed consuming bananas and coconuts according to the model above. Observing the trajectory of consumption of bananas, we would find a dynamic signal of the form $Q_B(t)=j_B\left(1-e^{-r_B\cdot t}\right)$. Likewise, we would find a tsignal for coconuts of the form $Q_C(t)=j_C\left(e^{-r_B\cdot t}-1\right)$. From these parametric forms, which we have derived from, say, observing the diminishing total number of bananas and coconuts on Robinson Crusoe's island, we may surmise that these quantities are related, given the similarity of their forms, and suggest some constraint such that they linearly sum up to some constant (i.e. a quantity we construct as the budget constraint). 

We also notice the trajectory of bananas are of a frictional sort by computing the force on that signal and realizing, in trying to construct a Lagrangian, that this force is not exactly integrable. That frictional force, incidentally, resembles that of a viscous medium, and the trajectory resembles that of an object moving at constant speed entering a viscous medium gradually coming to a stop. 

What we can say here is that whatever is driving the dynamic of ``consumption'' must be proportional to a non-increasing marginal quantity of some function. Indeed, rather than assuming \emph{a priori} non-increasing marginal utility (i.e. $\partial^2 U/\partial Q^2 \leq 0$) as the equilibrium-seeking force, we can empirically observe the force on the price signal, build a bespoke model of this force (such as by attributing it as the partial derivative of some function $U$ by $\partial Q$, and choose to interpret it (or some component of it) as due to non-increasing marginal utility on the part of a putative consumer.

The Lagrangian and Hamiltonian forms we may construct for representing the dynamic signals of a system will be free of the unit of utility, as discussed above. But we are also free of a necessary assumption that these signals are being driven by an agent or some collection of individual agents. There is only the notional representative agent, the system or market itself, generating these dynamic signals. But it is not clear that this representative agent is truly an agent with choice or that its signals are generated through the aggregation of individual agents' behaviors. Indeed, the microeconomic pedagogical elision between computing preferences for individual agents, then aggregating them up to some representative agent, itself a beast with its own mind, is not quite an agent with free will in the traditional sense. All we have are signals, to which we may devise a theory of agents with choices making decisions to explain how they are generated. Or, as the following section describes, we may simply choose an alternate non-agentic interpretation. In either case, the irony of a theory of choice and decision-making as representative of human agency (and for many, human freedom), reduces just as well to a non-agentic explanatory approach to a theory explaining signals.

\subsection{A short story of scientific iteration}

This alternate approach to modeling Robinson Crusoe's consumption in this ``market'' of bananas and coconuts can help us tell a brief story about how our scientific understanding could emerge and accommodate competing interpretations.

Imagine that we are on the mainland, and that we have been maintaining a careful watch over a nearby island through our telescope. Every day, we count the bananas and coconuts on the thickly forested island, maintaining a plot of counts over time, and we are reasonably certain our counts are comprehensive. One day we notice a significant deviation in trends of our counts, suggesting to us that something unusual is happening. We find that bananas and coconuts are declining at certain rates, and with the rate of decline gradually diminishing. Using the approach described above, we have used these counts as a signal, and from them we have constructed a Lagrangian/Hamiltonian that fits the respective Euler-Lagrange and Hamilton-Jacobi equations. From the signals we find that the changes in each signal are collinear. We also find that some drag force seems to be acting on their velocity. What are we to make of these signals?

At our weekly mainland council, we field a couple of theories. One proposed theory is that the bananas and coconuts have begun to wash away with the tides. The reasoning here is that the island had reached some peak capacity for bananas and coconuts, and that this abundance of tree fruits has started to push some toward the beach to be washed away by the tides. We, the council members, nod our heads in vigorous agreement, but you are unsure. What about the drag force, the diminishing rate of decline? The going theory proposes that as the tide washes away some fruit, there are fewer fruits to jostle others to the edge of the beach. While that seems reasonable, you still wonder: why would the change in observables of bananas and coconuts be colinear if they were both at equal risk to be washed away by the tide? Your doubt leads you to suggest that the council look for bananas and coconuts accumulating on the beaches through the telescope, and as you suspect, the beaches appear clear of these fruits.

You dwell on the collinearity of the signals of bananas and coconuts, and the declining rate of decrease over time. While at the docks, you had heard of an enterprising sailor, Robinson Crusoe, who had left shortly before the signals exhibited this perturbation. Well, you expound to the council, Robinson Crusoe had perhaps washed up on the island, and had started to collect bananas and coconuts for sustenance. He is, however, not an equal opportunist consumer of bananas and coconuts. He prefers one to the other, enjoys variety, and has a diminishing marginal utility for the consumption of each as well as a fixed energy budget to pick and consume them. This theory recommends itself to the council because it also matches the signals and their corresponding Lagrangian/Hamiltonian, while also taking the collinearity of the signals' velocities into consideration. 

But if this was the case, what of this Robinson Crusoe fellow? You suggest a means of confirmation: add an additional scope lens to your telescope. Perhaps Robinson Crusoe has been concealed by the thick foliage of the island, and a closer look would help you discover if he was stranded there. And lo and behold, there he is, hiding in the branches eating bananas and coconuts. The council swiftly sends an expedition to the island to rescue him.

This fanciful story illustrates two key point. One is that our signals-based approach can indeed lead to cumulative scientific discourse blending empirical observation and model-building, without having to adopt a specific \emph{ex ante} theoretical framework to drive it forward. We can, in a sense, embed an agnostic approach to theory-development into the mathematical recipe for modeling itself. Second, our approach can be complementary to the use of rational-agent, utility-maximizing, preferences-having models as a mode of explanation, as it does in this story. The difference is an agnosticism that leaves open the possibility of alternative explanations and theories that may be more suitable given the observed signals. Again, the point is not to reject the traditional approach to microeconomic modeling out-of-hand, but to see it as one way to theorize and model compatible with a more generalized epistemically-grounded approach to scientific inquiry that accounts for time.

\section{Conclusion}

What makes for a good economic theory? Economic theories are often held out to abide by a set of additional standards by which the community of economic scholars judge their own theories. Drawn in many ways from their colleagues in physics, economists evaluate theories by parsimony, internally consistency, descriptive (and prescriptive) strength, and the ability to place theory in dialogue with empirical observation.

The signals-based approach outlined here, especially using Lagrangian and Hamiltonian dynamics, fits these above criteria. Critics may argue, that with utility functions, price functions, etc. of greater complexity than the ones I have used here, the development of parsimonious model forms becomes increasingly untenable, but this remains true even in traditional choice and decision theories as we add more agents, aggregation functions, and strategic considerations through game theory. All parsimonious heuristic models must eventually be content with a far more complex and less parsimonious form in order to make useful contact with empirical observations. Furthermore, I believe our approach, in focusing first on the signal itself rather than underlying preferences or utility function, offers a far clearer comparison with empirical observations over time.

It is even clearer where the theory is meant to be descriptive versus prescriptive. In traditional economic theory, the normative framework that preference satisfaction is \emph{per se} good encourages a slippage between a description of a market system and a prescription of how a market system ought to be organized. Here, in this approach, I adhere strictly to descriptive models of observed signals, and leave the evaluative framework over those signals to normative theorists to devise and advocate for themselves.

I hope that this alternative approach to formalization is a bridge, not a destination, toward whatever comes next. If the formal language itself can encompass its own epistemic agnosticism and uncertainty, if the mechanical workings of the model itself reflect important features of our observations, such as the passage of time, emergent structures, and the perturbation of information, and if our model-making can make better contact with narrative storytelling, qualitative research, and other means of sparking our curiosity for understanding beyond the veil of the model, then we have found our way to a better climate even if we have yet to reach Shangri-La.


\section*{Acknowledgments}
The author would like to thank the organizers and participants of the Economic Methodology: Models, Measurements, and Interventions Conference 2024, Steve Vogel and the participants of the Society for the Advancement of Socio-Economics Conference 2024, Patrick Klösel, Lucas Osborne, Luke Herrine, and David Grewal for helpful feedback, commentary, and support in the preparation of this article.

\section*{Disclosure of interest}
The author reports there are no competing interests to declare.

\section*{Biographical Note}
Sarang Shah is a PhD candidate in Political Science at the University of California, Berkeley.

\bibliographystyle{apsr}
\bibliography{bibliography}

\appendix

\section{Robinson Crusoe Model of Exchange}\label{appendix}

In this section, we examine the construction of dynamics in the context of exchange.

Robinson Crusoe has washed up on a desert island, except this time he is accompanied by his first mate, Defoe. Robinson Crusoe and Defoe have a treacherous relationship, and immediately upon crashing on the island, they each retreat to opposing corners. There, they each begin collecting some endowment of the bananas and coconuts on the island until they have together collected all of them. Alas, Robinson Crusoe and Defoe both realize that their endowments are not quite to their liking, and that by trading with each other, they could each get a better distribution of bananas and coconuts.

The principles of exchange are simple. For stylistic simplicity, let us assume that Robinson Crusoe has far too great a proportion of bananas to coconuts, and Defoe has far too great a proportion of coconuts to bananas. Robinson Crusoe wishes to lower that proportion by trading bananas for more coconuts, and likewise, Defoe wishes to trade coconuts to bananas. There is some ratio $r$ the parties agree upon for trading bananas for coconuts (e.g. 3 bananas for 2 coconuts). This ratio effectively functions as price in this scenario, with Robinson Crusoe's endowment of bananas serving as his total "budget", and Defoe's coconuts serving as his. Each seeks to spend down their respective budget to achieve higher utility. The budget for each can be expressed as $Q^{RC}_B = rQ^{RC}_C$ and $Q^{D}_B = rQ^{D}_C$.

We presume that neither trades when the terms of the trade (e.g. $r$) would put them in a position where either has lowered their utility. Naturally, as Robinson Crusoe hands over bananas to Defoe, Defoe's accumulation of bananas increases while Robinson Crusoe's decreases. The same goes for coconuts. 

As with the previous example, we assume Robinson Crusoe and Defoe each have their own utility function. For Robinson Crusoe, $U^{RC} = kQ^{RC}_BQ^{RC}_C$ and for Defoe, $U^{D} = lQ^{D}_BQ^{D}_C$. Finally, we will start by assuming that $r$ is constant. As will be shown later, and shown in \cite{Glotzl_Glotzl_Richters_2019}, $r$ must be within certain tangential limits in order for exchange to proceed, and that $r$ would have to be at the \cite{Arrow_Hahn_1971} first-order equilibrium conditions in order to maximize, rather than simply Pareto-optimize, the utilities of both parties. For now, it suffices to take the Walrasian auctioneer as someone who wishes above all for price to be stable, thus fixed. It is up to the two parties, Robinson Crusoe and Defoe, to determine how much they wish to trade with each other at any given time to continually optimize their respective utilities.

\subsubsection{Static Exchange}

\cite{Arrow_Hahn_1971} succinctly give us the equilibrium condition for each party. For Robinson Crusoe, equilibrium is found when:
\begin{equation}
\begin{split}
r\frac{\partial U^{RC}}{\partial Q^{RC}_B} &= \frac{\partial U^{RC}}{\partial Q^{RC}_C} \\
rkQ^{RC}_C &= kQ^{RC}_B \\
r &= Q^{RC}_B/Q^{RC}_C
\end{split}
\end{equation}
The total endowment remains the same before and after exchange. So while holding that $Q^{RC}_B+Q^{D}_B=Q^{T}_B$ and $Q^{RC}_C+Q^{D}_C=Q^{T}_C$, we can also find the quantities possessed by Defoe at equilibrium. What this means is that possible equilibria of exchange lie on a line passing through the endowment with slope $-r$ in an Edgeworth Box description of quantities possessed by each party.

\subsubsection{Dynamic Exchange}

The relationship of exchange in the dynamic view can be represented as follows. As Robinson Crusoe exchanges bananas for coconuts with Defoe, the magnitude of Defoe's rate of increase of bananas is equal to that of Robinson Crusoe's decrease. Likewise, Robinson Crusoe's coconuts increase at the same rate as Defoe's coconuts decrease:

\begin{equation}
\dot{Q}^{RC}_B = - \dot{Q}^{D}_B
\end{equation}
\begin{equation}
\dot{Q}^{RC}_C = - \dot{Q}^{D}_C
\end{equation}

This is equivalent to the finding that the total endowment remains unchanged after exchange.

The exchange itself can be expressed as $\dot{Q}^{RC}_B = -r\dot{Q}^{RC}_C$, i.e. Robinson Crusoe trading bananas at price $r$ yields to an $r$ proportioned increase in coconuts.

Let us now solve for the dynamic trajectory of Robinson Crusoe's stash of coconuts. As with the previous example, we start by positing a force on Robinson Crusoe's acquisition of coconuts as proportional to the marginal utility he gains from exchanging bananas for coconuts:
\begin{equation}
\begin{split}
f^{RC}_C &= \partial U^{RC}/\partial Q^{RC}_C\\
&= \partial/\partial Q^{RC}_C\left(kQ^{RC}_BQ^{RC}_C\right)
\end{split}
\end{equation}

Before we proceed further, recall the exchange equation $\dot{Q^{RC}_B} = -r\dot{Q^{RC}_C}$. Integrating both sides with the objective of solving for $Q^{RC}_B$ yields:
\begin{equation}
\begin{split}
\int_0^t \dot{Q}^{RC}_B &= -r\int_0^t \dot{Q}^{RC}_C \\
Q^{RC}_B(t)-Q^{RC}_B(0) &= -r\left[Q^{RC}_C(t)-Q^{RC}_C(0)\right]\\
Q^{RC}_B(t) &= -rQ^{RC}_C(t)+r\left[Q^{RC}_B(0)+Q^{RC}_C(0)\right]
\end{split}
\end{equation}

Simplifying our notation by defining $\theta^{RC} \equiv Q^{RC}_C(0)+rQ^{RC}_B(0)$, we find $Q^{RC}_B(t)=\theta^{RC}-rQ^{RC}_C(t)$. Plugging this into our force equation, we find:

\begin{equation}
\begin{split}
f^{RC}_C &= \partial/\partial Q^{RC}_C\left(kQ^{RC}_BQ^{RC}_C\right) \\
&= k\partial/\partial Q^{RC}_C\left(\left(\theta^{RC}-rQ^{RC}_C\right)Q^{RC}_C\right) \\
&= k\left[\theta^{RC}-2rQ^{RC}_C\right]
\end{split}
\end{equation}

At this stage, observant readers may notice the similarity of this expression with the force in our previous consumption example. This is no accident, if we think of exchange as consumption along some path prescribed by a fixed $r$ as price. Of course, that $r$ must be such that it falls between the tangent slopes of each party's indifference curves at the initial endowment, otherwise these parties would see little reason to participate in exchange. Nevertheless, given a suitable $r$, our formula of taking marginal utility as proportional to force will lead us down a Pareto-optimal outcome. Given that we have arrived at a similar functional form as the simple consumption case, with the budget term $E$ replaced by $\theta$, we will spare the reader a familiar elaboration of the differential equation solution to $f = m\dot{Q}$ and the corresponding Lagrangian and Hamiltonian formulation.

This may be unsatisfying to readers, with whom we are inclined to agree. Fortunately, the excellent work of \cite{Glotzl_Glotzl_Richters_2019} explores the possible dynamic implications of a non-fixed $r$ for three different sorts of Walrasian auctioneers and \emph{tâtonnement}, with an indeterminate utility function for each party. In each of these cases, it is still possible to compute some path-dependent trajectory toward a Pareto-optimal outcome. We chose this exceedingly simple example to show the basics of exchange are susceptible to a dynamic approach, while highlighting the key features of this approach that could be applied to any manner of utility functions or auctioneers. 

While some may object that the key feature of exchange is indeed price-setting, the auctioneer, and groping toward the outcome that maximizes the sum of utilities, we would like to note the following. The work of the auctioneer, or the work of negotiation, is an additional dynamic term that must be introduced exogenously into a dynamic understanding of exchange. This term, that we would call $r(t)$ possesses two qualities. First, it is bounded by the tangent lines to the indifference curves at the initial point of endowment (and indeed at any point of exchange). The slope $r(t)$ is always set within the range of the tangent lines of ever smaller \emph{vesicae piscis} formed by the indifference curves of each party. Second, without knowing how much the other party possesses, it is impossible for each party to determine the optimal $r$ for both parties. All they have is the internal compass, in this case, of whether exchange will or will not marginally increase utility. A third party however could collect information on both parties and set this rate. Finally, prices are often the result of many factors, not just a rational auctioneering or negotiation process. By setting aside $r(t)$ as a dynamic term worthy of explanation itself, from which and through which the other dynamic $Q(t)$ terms may be integrated, allows us to focus our attention on what exactly influences price. In fact, we can even discern the path of $r(t)$ from observation of the $Q(t)$ themselves. In any case, the point of this exercise is that even exchange can be framed in a dynamic manner, and that in doing so, we can investigate interesting features of the dynamic of exchange that would have otherwise remained obscured.
\end{document}